\documentclass[prl,showpacs,twocolumn]{revtex4}
\usepackage{graphicx}
\usepackage{bm}
\usepackage{pifont}
\usepackage{amsmath}
\begin{document}
\title{Pseudogap of metallic layered nickelate $R_{2-x}$Sr$_x$NiO$_4$ ($R=$ Nd, Eu) crystals measured using angle-resolved photoemission spectroscopy}
\author{
M. Uchida$^{1}$, K. Ishizaka$^1$, P. Hansmann$^{2}$, Y. Kaneko$^3$, Y. Ishida$^4$, X. Yang$^5$, R. Kumai$^6$, A. Toschi$^2$, Y. Onose$^{1,3}$, R. Arita$^1$, K. Held$^{2}$, O. K. Andersen$^{5}$, S. Shin$^{4,7}$ and Y. Tokura$^{1,3,8}$
}
\affiliation{
$^1$Department of Applied Physics, University of Tokyo, Tokyo 113-8656, Japan \\ $^2$Institut for Solid State Physics, Vienna University of Technology, 1040 Vienna, Austria \\ $^3$Multiferroics Project, ERATO, Japan Science and Technology Agency (JST), Tokyo 113-8656, Japan \\ $^4$Institute for Solid State Physics (ISSP), University of Tokyo, Chiba 277-8561, Japan \\ $^5$Max-Planck-Institut f\"{u}r Festk\"{o}rperforschung, D-70569 Stuttgart, Germany \\ $^6$National Institute of Advanced Industrial Science and Technology (AIST), Tsukuba 305-8562, Japan \\ $^7$RIKEN SPring-8 Center, Hyogo 679-5148, Japan \\ $^8$Cross-Correlated Materials Research Group (CMRG) and Correlated Electron Research Group (CERG), ASI, RIKEN, Wako 351-0198, Japan
}
\begin{abstract}
We have investigated charge dynamics and electronic structures for single crystals of metallic layered nickelates, $R_{2-x}$Sr$_x$NiO$_4$ ($R=$ Nd, Eu),
isostructural to La$_{2-x}$Sr$_x$CuO$_4$.
Angle-resolved photoemission spectroscopy on the barely-metallic Eu$_{0.9}$Sr$_{1.1}$NiO$_4$ ($R=\mathrm{Eu}$, $x=1.1$) has revealed
a large hole surface of $x^2-y^2$ character with a high-energy pseudogap of the same symmetry and comparable magnitude with those of underdoped ($x<0.1$) cuprates,
although the antiferromagnetic interactions are one order of magnitude smaller.
This finding strongly indicates that the momentum-dependent pseudogap feature in the layered nickelate arises from the real-space charge correlation.
\end{abstract}
\maketitle

\begin{figure}
\begin{center}
\includegraphics*[width=7.8cm]{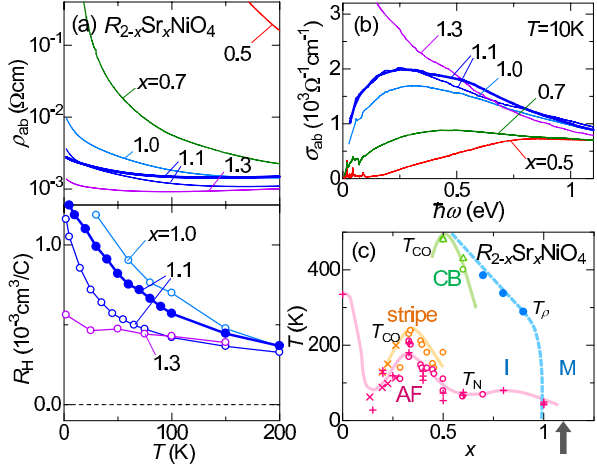}
\caption{(color online).
(a) Temperature dependence of the in-plane resistivity (upper panel) and Hall coefficient (lower panel).
(b) Doping variation of the in-plane optical conductivity spectra at 10 K,
as deduced by Kramers-Kronig transformation of the respective reflectivity data in the photon energy range of 0.01-40 eV.
Thin lines show the results in an wide doping range for $R=\mathrm{Nd}$ and bold lines for $R=\mathrm{Eu}$.
(c) Doping phase diagram for $R_{2-x}$Sr$_x$NiO$_4$ as reproduced from the literature
($\bigcirc $ \cite{Nineutron}, $\bigtriangleup$ \cite{Nisigma}, $\times$ \cite{Nineutrontra}, and $+$ \cite{Nimsr}).
The parent compound ($x=0$) is an antiferromagnetic (AF) insulator with $T_{\mathrm{N}} \sim 330$ K.
By hole doping, it shows diagonal-stripe and checkerboard (CB) type charge ordering (CO) centered at $x=1/3$ and $x=1/2$, respectively,
and the ground state remains insulating up to $x\sim 1$.
Filled circles indicate the resistivity upturn temperature $T_{\rho}$ toward the resistivity divergence (insulator state)
below $x=0.9$
and a dashed line represents the approximate position of the insulator (I)-metal (M) phase boundary.
}
\label{fig1}
\end{center}
\end{figure}

High-$T_{\mathrm{c}}$ superconductivity appears close to the Mott transition induced by doping holes or electrons into antiferromagnetic parent insulators.
Such filling-control insulator-metal transitions are widely observed for transition-metal oxides with correlated electrons \cite{MIT},
yet the emergence of high-$T_{\mathrm{c}}$ superconductivity remains unique for the layered cuprates.
Layered nickelate $R_{2-x}$Sr$_x$NiO$_4$ (RSNO, $R$ being rare earth element) with K$_2$NiF$_4$ type structure
is a rare example of a two-dimensional antiferromagnetic insulator-metal transition system \cite{MIT},
providing a contrastive counterpart to superconducting La$_{2-x}$Sr$_x$CuO$_4$ (LSCO) with the same crystal structure.
In RSNO, diagonal-stripe charge ordering is observed at $x\sim 1/3$ \cite{Nistripedoping, Nistripeele}
in contrast to LSCO with vertical-stripe charge ordering at $x\sim 1/8$ \cite{firstCustripe}.
RSNO also shows checkerboard charge ordering at $x\sim 1/2$ \cite{Nineutron, Nistripeele}.
This system undergoes an insulator-metal transition with melting of its charge ordered state
but shows no superconductivity up to $x=1.6$ \cite{Nipoly}.
Hitherto there has been no study of the electronic band structure in the insulator-metal critical region.
The comparison between these layered nickelate and cuprate Mott transition systems will be of benefit for understanding the uniqueness of high-$T_{\mathrm{c}}$ cuprates.

We have succeeded in growing single crystals of RSNO up to $x=1.3$, i.e. \textit{metallic} crystals,
by optimizing the $R$ species and using a high gas-pressure floating zone method. 
Figure 1(a) summarizes the transport properties:
The in-plane resistivity decreases with increasing doping and
eventually shows no divergence in the zero temperature ($T$) limit above $x=1.0$,
signaling the emergence of the metallic ground state therein.
The weak upturn of the resistivity above $x=1.0$ was reported to be due to the weak localization \cite{Nithinfilm},
and we also confirmed that our result on the conductivity $\sigma \equiv 1/\rho$ well fits the relationship;
$\sigma =\sigma _0 +AT^{1/2}$ where $\sigma _0 = 260/\Omega \mathrm{cm}$ and $A= 92/\Omega \mathrm{cmK}^{1/2}$ for $x=1.1$.
The in-plane Hall resistivity below $x=1.1$ increases monotonically with lowering temperature.
This is analogous to the behavior observed in the underdoped cuprates,
as discussed in terms of pseudogap phenomena and/or antiferromagnetic correlation \cite{pgreview}.
In contrast, the Hall resistivity shows a fairly weak temperature dependence at $x=1.3$, as in a conventional normal (or overdoped) metal.
In Fig. 1(b), we show the doping variation of the in-plane optical conductivity spectra for RSNO ($R=\mathrm{Nd}$, Eu) at 10 K.
A pseudogap structure of about 0.2 eV clearly exists even in the metallic region at $x= 1.0$-1.1 and it evolves into the Drude peak at $x=1.3$.
This doping evolution is similar to the previous results for La$_{2-x}$Sr$_x$NiO$_4$ thin films \cite{Nithinfilm}.
These transport and optical properties indicate that an anomalous metallic state is realized at $x\sim 1.1$
as contrasted to the normal metallic behavior above $x\sim 1.3$.
Here we report angle-resolved photoemission spectroscopy (ARPES) experiments for the metallic layered nickelate Eu$_{0.9}$Sr$_{1.1}$NiO$_4$ (ESNO),
which is located close to the insulator-metal boundary shown by a gray arrow in Fig. 1(c).

Single crystals of RSNO ($R=\mathrm{Nd}$, Eu) were
grown with the floating-zone method in a high-pressure oxygen atmosphere ($p_{\mathrm{O}_2}=10$-60 atm).
The crystals obtained have a typical dimension of $2\times 2\times 0.5$ mm$^3$ for $x=1.1$.
In RSNO systems, the electronic phase diagram was reported to be least affected by variation of $R$ species \cite{Nisigma, Nineutron,Niorthostripe},
which is also confirmed up to a higher $x$ region in the present study as shown in Figs. 1(a) and (b).
We chose $R=\mathrm{Eu}$ (ESNO) for ARPES experiments because of its better sample quality and cleavage property.
Laser-ARPES measurements with an excitation energy $h\nu =6.994$ eV and an instrumental energy resolution of 3 meV
were carried out at ISSP, University of Tokyo using a VG-Scienta R4000 electron analyzer \cite{machine}.
Unless otherwise noted, the measurements were performed with circular polarized light to observe all the orbital components.
The samples were cleaved and measured at 7 K and the pressure was below $2\times 10^{-11}$ Torr throughout the measurement.
We confirmed that the sample degradation was negligible during the measurement.
The local-density approximation (LDA) calculation for the electronic structure of ESNO was performed with the LMTO and NMTO methods \cite{LMTO}.
In order to avoid the LDA for the $4f$ electrons in Eu$^{3+}$, they were treated as frozen core.
This makes it inconvenient to use the virtual-crystal approximation for Eu$_{0.9}$Sr$_{1.1}$NiO$_{4}$,
which was therefore substituted by Eu(Rb$_{0.1}$Sr$_{0.9}$)NiO$_{4}$
with the virtual Rb$_{0.1}$Sr$_{0.9}$-atom, the one with 0.1 proton less than Sr.

\begin{figure}
\begin{center}
\includegraphics*[width=8.4cm]{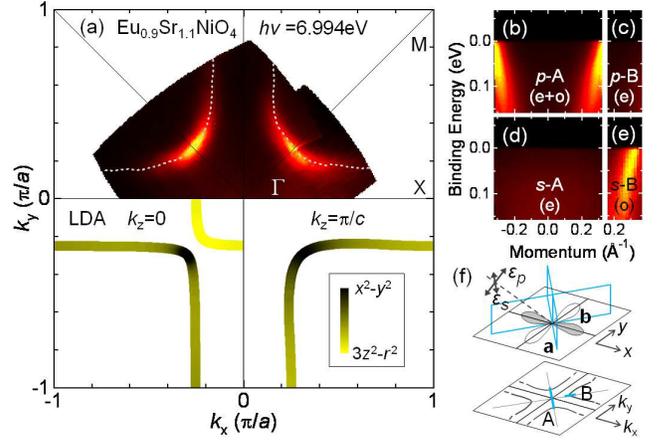}
\caption{(color online).
(a) Fermi surface of Eu$_{0.9}$Sr$_{1.1}$NiO$_4$ determined by integrating the spectral weight in a $\pm$ 4 meV window around $E_{\mathrm{F}}$.
The gray dashed  lines show the $k_{\mathrm{F}}$ positions determined by the momentum distribution curves (MDCs) at $E_{\mathrm{F}}$.
We set the work function at $\phi =4.0$ eV and confirmed that its variation causes no difference in our conclusion.
The lower panel shows the LDA calculation results.
(b)-(e) Image plots with $p$ and $s$ linear polarization at the cuts A and B shown in (f).
(f) Geometry of the light polarization vectors, mirror planes and the corresponding cuts in real and momentum space.
Each polarization has odd (o) or even (e) symmetry for the mirror planes (\textbf{a} and \textbf{b}) and the corresponding cuts (A and B) as shown in (b)-(e).
}
\label{fig2}
\end{center}
\end{figure}

In the upper half of Fig. 2(a), we show the Fermi surface for ESNO as revealed by ARPES.
Only one sheet of the Fermi surface is discerned, which is a large hole-like one centered at the M point.
To determine its orbital character,
we measured the polarization-dependent ARPES in the experimental configurations shown in Fig. 2(f).
The results (Figs. 2(b)-(e)) clearly show that the band dispersion along the momentum cut A (B) appears only in $p$ ($s$) configuration.
From the matrix element argument \cite{selectionrule}, the corresponding wave function should be characterized by
odd parity with respect to the mirror planes \textbf{a} ($110$) and \textbf{b} ($1\bar{1} 0$) shown in Fig. 2 (f).
This indicates that the large hole sheet has $x^{2}-y^{2}$ character at the point where it crosses the $\Gamma$-M line and
suggests that the low-energy physics of ESNO is dominated by the antibonding $pd\sigma$ O$_{x}$-Ni$_{x^{2}-y^{2}}$-O$_{y}$ orbital,
like that of the high-$T_{\mathrm{c}}$ cuprates.
This is not trivial, because one expects not only $x^{2}-y^{2}$ but also $3z^{2}-r^2$ character when the $e_{g}$ shell is 1/4 rather than 3/4 filled.
However, hybridization between $x^{2}-y^{2}$ orbital and any axial orbital such as $3z^{2}-r^2$ is forbidden along the $\Gamma$-M line \cite{OKA}.
As a consequence, at a $\Gamma$-M line crossing, the Fermi surface must have either $x^{2}-y^{2}$ or $3z^{2}-r^2$ character.
But there can be several crossings, and this is in fact what happens in the lower half of Fig. 2(a),
where we show the cross sections with the $k_{z}=0$ and $k_{z}=\pi/c$ planes of the Fermi surface of ESNO calculated in LDA.
The black/yellow coloring (dark/light contrast) gives the relative $\left( x^{2}-y^{2}\right) /\left( 3z^{2}-r^2 \right) $ $e_{g}$-Wannier-orbital character.
Apart from the large, predominantly $\left( x^{2}-y^{2}\right) $-like,
cylindrical hole sheet which resembles the one measured by ARPES,
the LDA gives a small, $\left(  3z^{2}-r^2 \right)  $-like $\Gamma$-centered electron pocket.
This LDA Fermi surface is very much like the one calculated for the LaNiO$_{3}$/LaAlO$_{3}$ heterostructure \cite{PH} and
we expect that, like in that case, the electronic Coulomb correlations treated properly
will remove the small electron pocket through enhancement of the $e_{g}$ crystal-field splitting.
At the moment, however, we cannot rule out the possibility that we are accidentally observing the specific $k_z$ plane with no electron pocket.
For further investigations of the $k_z$-dependent electronic structure, $h\nu$-dependent ARPES experiment is highly desired.

\begin{figure}
\begin{center}
\includegraphics*[width=8.4cm]{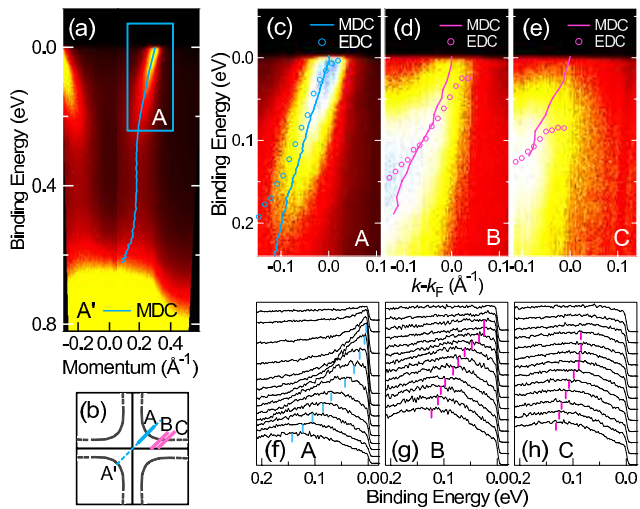}
\caption{(color online).
(a) ARPES spectra of Eu$_{0.9}$Sr$_{1.1}$NiO$_4$ along the $\Gamma$-M direction over a wide energy and momentum region,
showing a kink around 0.25 eV.
(c)-(h) Image plots and stacks of energy distribution curve (EDC) corresponding to the different cuts as sketched in (b).
Also shown on the plots (a), (c)-(e) are the dispersions obtained by following the peak positions of the MDC (solid lines) and the EDC (circles).
}
\label{fig3}
\end{center}
\end{figure}

As shown in Fig. 3(a), the band shows a symmetric dispersion around the $\Gamma$ point up to 0.6 eV along the $\Gamma$-M nodal direction.
Hereafter, we use the term `nodal' also for the present result on the nickelates,
to stress the clear resemblance of the pseudogap phenomena common to these materials, as will be shown later.
One can clearly see a kink-like structure around 0.25 eV, which is very similar to the so-called giant kink observed in cuprates \cite{firstgiant, hierar, extrinsic}.
The characteristic energy scale is 0.3-0.6 eV in the cuprates
and a number of origins have been suggested, including the effects of coupling to some bosons,
strong electron correlation \cite{hierar}, and photoemission matrix elements \cite{extrinsic}.
Recently, broad but similar features have been observed around 0.2 eV
also in other correlated oxides such as LaNiO$_3$ \cite{LNO} and SrVO$_3$ \cite{SVO}.
Therefore, this high-energy anomaly seems to be a momentum-resolved ubiquitous feature observed in Mott transition systems,
and the convincing theory must explain this ubiquity and the difference of their energy scales.
As shown in Fig. 3(c), on the other hand, ESNO does not clearly show any kink structure at $\sim 70$ meV,
which has been universally observed
among the various layered cuprates \cite{firstkinkone, firstkinktwo, dopingkinkone} and discussed in the light of the electron coupling to a bosonic mode
(phonon \cite{dopingkinkone} or spin excitation \cite{firstkinktwo, dopingkinktwo})
as a possible glue for the Cooper pair.
The absence of the low-energy kink in non-superconducting ESNO conversely suggests its possible key role for the emergence of high-$T_{\mathrm{c}}$ superconductivity
unique in the layered cuprates.

Hereafter, we focus on the near-$E_{\mathrm{F}}$ electronic structure along off-nodal directions.
Figures 3(d) and (e) show the ARPES image plots along the momentum cuts B and C, as sketched in Fig. 3(b).
On approaching the ($\pi$, 0) region, the clear quasiparticle dispersion gets suppressed and a gap-like feature appears at $E_{\mathrm{F}}$.
We can clearly see this nodal-antinodal dichotomy in the energy distribution curves (EDCs) shown in Figs. 3(f)-(h).
While the sharp quasiparticle peak is found close to the nodal points,
the spectral weight around $E_{\mathrm{F}}$ becomes largely suppressed upon moving away from the nodal region.
This feature demonstrates a marked resemblance to the high-energy pseudogap
with the same momentum-space symmetry as the $d$-wave superconducting gap \cite{firstsmallpg, pgreview, largepg}.
It is universally observed for underdoped cuprates and believed to be an intrinsic feature of the high-$T_{\mathrm{c}}$ superconductors.

\begin{figure}
\begin{center}
\includegraphics*[width=6.9cm]{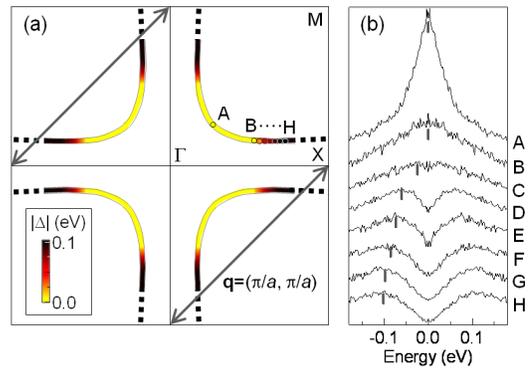}
\caption{(color online).
(a) Momentum dependence of the pseudogap in Eu$_{0.9}$Sr$_{1.1}$NiO$_4$.
The pseudogap size $|\Delta|$ is estimated from the peak position of the symmetrized EDC spectra shown in (b).
The open circles labeled with alphabets (A to H) denote the $k_{\mathrm{F}}$ positions.
The gray arrows indicate the wave vector of the expected charge correlation in this system.
(b) A sharp nodal quasiparticle peak is transformed into a broad pseudogap feature over $\sim0.1$ eV while approaching the ($\pi$, 0) region.
}
\label{fig4}
\end{center}
\end{figure}

In Fig. 4(a), we show the momentum-dependent pseudogap energy along the expected Fermi surface,
which is estimated by the peak position of the symmetrized EDC spectra at $k_{\mathrm{F}}$ as shown in Fig. 4(b).
This behavior is strikingly similar to the Fermi-arc phenomena in the cuprates \cite{largepg}.
On approaching the ($\pi$, 0) region, the sharp quasiparticle peak is suppressed and converted into a broad gapped feature over $\sim 0.1$ eV.
Twice the value of its energy scale ($\sim$ 0.2 eV)
is well in accord with that of the pseudogap observed in the optical conductivity spectra (Fig. 1(b)).
Its consistency indicates that the pseudogap in optical spectroscopy reflects the structure around ($\pi$, 0).
This Fermi-arc like behavior also reflects the transport properties.
In case no pseudogap is considered on the Fermi surface,
we can estimate the Hall coefficient to be $R_{\mathrm{H}}^{\mathrm{ARPES}} =4\pm 1\times 10^{-4} \mathrm{cm}^3/\mathrm{C}$
from the area of the Fermi surface.
It approximately corresponds to the high-temperature ($\sim$ 200 K) $R_{\mathrm{H}}$ value,
and its increase with lowering temperature can be interpreted as the opening of the pseudogap (Fig. 1(a)).
Thus, the momentum-dependent pseudogap is an intrinsic feature
which also dominates the charge dynamics near the insulator-metal transition in this system.

Many theoretical models have been proposed for explaining the origin of the pseudogap so far \cite{pgreview}.
The antiferromagnetic interaction in the nickelate parent material is one order of magnitude smaller than that of cuprates \cite{NiJ}.
Considering the comparable size of the high-energy pseudogap, it is unlikely that its origin is related to the spin correlation.
In the layered nickelates, the checkerboard type charge ordering persists above $x=0.5$ with introducing the excess holes to $d_{3z^2-r^2}$ orbital states,
and its melting plays a key role in driving the insulator-metal transition around $x\sim 1$ \cite{Nithinfilm, Nisigma}.
Therefore, the high-energy checkerboard type charge correlation should give rise to the momentum-dependent pseudogap,
which may particularly affect the ($\pi$, 0) and equivalent regions connected with its wave vector $\textbf{q} =(\pi /a, \pi /a)$.
Also in the layered cuprates, various charge ordering or correlation patterns have been reported so far.
The vertical-stripe type charge ordering is observed by inelastic neutron scattering in LSCO \cite{firstCustripe},
whereas $4a\times 4a$ checkerboard type charge ordering is observed by scanning tunneling microscopy
in Ca$_{2-x}$Na$_x$CuO$_2$Cl$_2$ \cite{COCNCOC}.
In both cases, a strong dichotomy of the quasiparticle peak at nodal and antinodal regions is observed by ARPES \cite{pgCNCOC}.
Here it is worth noting that similar dichotomy is observed in a ferromagnetic bilayer manganite
with local charge-ordering correlations \cite{pgMn}.
All these results may suggest the possibility that the Fermi surface dichotomy with the high-energy pseudogap
commonly arises from the real-space charge correlation.

Our ARPES experiments thus show that the metallic states in layered nickelates and cuprates share the basic electronic structures
such as the $d_{x^2-y^2}$ orbital-derived single Fermi surface with similar shape
and Fermi-arc dichotomy with the high-energy pseudogap at the antinodal region.
The high-energy pseudogap may be commonly interpreted as the manifestation of the real-space charge correlation.
On the other hand, the clear difference appears in low-energy ($<100$ meV) features,
as represented by the lack of the low-energy kink and small pseudogaps.
Therefore, finding the origin of these similarities and differences may contribute to further understanding of the microscopic mechanism
behind the high-$T_{\mathrm{c}}$ superconductivity.

We thank N. Nagaosa, J. Fujioka and S. Seki for fruitful discussions.
This work was partially supported by Grants-in-Aid for Scientific Research (Grants No. 20046004 and No. 20340086) from the MEXT of Japan and JSPS
and by Funding Program for World-Leading Innovative R \& D on Science and Technology (FIRST Program).
M.U. acknowledges the support by a Grant-in-Aid for JSPS Fellows (Grant No. 21-5941).
K.H. and O.K.A. appreciate support by the EU network MONAMI.
P.H. has been supported by the FWF OK W004.

\end{document}